\documentclass[conference]{IEEEtran}

\usepackage{bbm}
\usepackage{amsmath, amsthm, amssymb}
\usepackage{graphicx}
\usepackage{pict2e,color}
\usepackage{multirow}
\usepackage{amsfonts}
\usepackage{verbatim} 
\usepackage{url}
\usepackage{empheq}
\usepackage[normalem]{ulem}
\usepackage{textcomp}
\newcommand{\figref}[1]{Figure \ref{#1}}

\usepackage{subcaption} 

\usepackage{cite}

\usepackage{algpseudocode}
\usepackage{algorithm}
\usepackage{tikz}
\usetikzlibrary{arrows}
\usepackage{tabularx}
\usepackage{dsfont}

\usepackage{caption}
\usepackage{subcaption}
\usepackage{dblfloatfix}

\usepackage{tikz}
\usetikzlibrary{shapes,arrows}

\usepackage{verbatim}

\tikzstyle{decision} = [diamond, draw, fill=blue!20, 
    text width=4.5em, text badly centered, node distance=3cm, inner sep=0pt]
\tikzstyle{block} = [rectangle, draw, fill=blue!20, 
    text width=5.5em, text centered, rounded corners, minimum height=4em]
\tikzstyle{line} = [draw, -latex']
\tikzstyle{cloud} = [draw, ellipse,fill=red!20, node distance=3cm,
    minimum height=2em]

\usepackage[flushleft]{threeparttable}
\usepackage{subcaption}

\usepackage[numbers]{natbib}

\theoremstyle{definition}

\def\argmin{\mathop{\rm arg\,min}}%

\IEEEoverridecommandlockouts                              

\begin{document}

\title{Distribution System State Estimation in the Presence of High Solar Penetration}
\author{Thiagarajan Ramachandran, Andrew Reiman, Sai Pushpak Nandanoori, Mark Rice, and Soumya Kundu
	\thanks{T. Ramachandran, A. Reiman, S. Nandanoori, M. Rice and S. Kundu are with the Energy and Environment Directorate at Pacific Northwest National Laboratory, Richland, WA 99354. {\tt\small \{thiagarajan.ramachandran, andrew.reiman, saipushpak.n, mark.rice, soumya.kundu\}@pnnl.gov.}}
}

\maketitle

\begin{abstract}
Low-to-medium voltage distribution networks are experiencing rising levels of distributed energy resources, including renewable generation, along with improved sensing, communication, and automation infrastructure. As such, state estimation methods for distribution systems are becoming increasingly relevant as a means to enable better control strategies that can both leverage the benefits and mitigate the risks associated with high penetration of variable and uncertain distributed generation resources. The primary challenges of this problem include modeling complexities (nonlinear, nonconvex power-flow equations), limited availability of sensor measurements, and high penetration of uncertain renewable generation. This paper formulates the distribution system state estimation as a nonlinear, weighted, least squares problem, based on sensor measurements as well as forecast data (both load and generation). We investigate the sensitivity of state estimator accuracy to (load/generation) forecast uncertainties, sensor accuracy, and sensor coverage levels. 
\end{abstract}
\section{Introduction}
A combination of increasing concerns about the effect of emissions on the environment, aggressive state-level renewable portfolio standards, and decreasing cost of photovoltaic (PV) solar panels are spurring adoption of solar generation in distribution networks. However, from an operational perspective, there are technical challenges associated with the introduction of variable distributed energy resources that need to be overcome to ensure the safety and  reliability of the distribution network. For instance, PV generation can cause over-voltage and can mask load from protection equipment. Furthermore, variation in cloud cover can cause voltage flicker and excessive operation of load tap changers \cite{seguin2016high,ari2011impact,cheng2016photovoltaic}.  These challenges can be addressed by advanced control strategies that benefit when distribution system operators have better \textit{observability} of the system state (i.e., the complex node voltages or branch currents).

Power system state estimation has been well studied from a transmission system perspective, and has been deployed in control centers for decades \cite{monticelli2012state,gomez2004power}. Distribution system state estimation (DSSE) is not as well established due to several modeling challenges. Distribution systems have lower reactance/resistance (X/R) ratios, which makes use of DC power flow approximations difficult. The unbalanced nature of the distribution system means that phase couplings need to be considered and the state estimation cannot be treated as a single-phase problem. Furthermore, distribution systems are not typically endowed with the adequate sensor coverage required for traditional state estimation algorithms to work. This is addressed in DSSE literature by introducing \textit{pseudo-measurements} with high variance, which correspond to load injections at different nodes in the system, determined by historical load levels \cite{baldwin1993power,manitsas2012distribution}.

Static DSSE methods typically fall into three categories: 
\begin{enumerate}
    \item \textsc{Branch-current based DSSE}: Branch-current based DSSE methods treat branch currents as state variables and convert available sensor measurements and pseudo-measurements into current measurements, and then solve a nonlinear weighted least squares (WLS) problem for the most likely set of branch currents\cite{li2004branch}. These methods are designed specifically for distribution networks with unbalanced flows, and assume that the network topology is radial or weakly meshed. There have been several extensions to the branch-flow based methods that incorporate phasor measurement unit (PMU) measurements, address bad-data problems, and consider the effect of correlations between pseudo-measurements on the state estimate \cite{li2004branch,baran2009,muscas2014effects,pau2013efficient}.

\item \textsc{Voltage based DSSE}: Voltage based state estimation methods take the node voltage magnitudes and phase angles as the system state and attempt to reconstruct the voltage phasors by computing the maximum likelihood estimate based on sensor measurements using a WLS approach \cite{baran1994state,lu1995distribution,chen1991distribution,li1996state}. The associated WLS problem is nonconvex in nature and makes no assumptions regarding the topology of the distribution network. A recent paper \cite{cruz2017two} on voltage based state estimation presents an approach to estimate the distribution system state with streaming measurements by developing an approach that updates the maximum likelihood estimate based on incoming measurements.

\item \textsc{Load allocation based methods}: The load allocation methods (\cite{dvzafic2013real}, \cite{roytelman2009real}) use the available measurements to construct/update an extended load forecast. Given the load forecast, a load flow problem is solved to estimate the system voltage profile. The method makes effective use of sparse measurements and has been field-tested in several distribution networks with real-time data. 
\end{enumerate}

This paper uses a WLS approach (similar to those detailed in \cite{baran1994state}-\cite{cruz2017two}) in order to set up and solve the state estimation problem for three different distribution feeders. The main contribution is twofold. First, we present a scenario construction approach that combines a load multiplier profile with solar generation data to generate net-load profiles typical of high PV penetration environments. Second, we present numerical results that show the sensitivity of the WLS approach to pseudo-measurement accuracy, measurement accuracy, and sensor coverage. We show that the relationship between the accuracy of the pseudo-measurements and the (less uncertain) sensor measurements is a key component in determining the accuracy of the state estimate.

The paper is structured as follows: Section \ref{sec:2} provides a description of the network and the measurement model. Section \ref{sec:3} consists of a brief description of the optimization problem. Section \ref{sec:4} details the scenario construction method, and Section \ref{sec:5} presents the numerical results.

\section{Network and Measurement Model}
\label{sec:2}
Consider a distribution network with $N$ buses. The state of the distribution system is given by the voltage vector $V \in \mathds{C}^M$.
The voltage vector is represented by $V$ and its dimension (denoted $M$) depends on the number of buses and the number of phases associated with each bus, which can range anywhere from 1 to 3 for distribution systems. Let $Y^{(M \times M)}$ represent the network admittance matrix associated with the distribution system $\mathcal{G}$ that satisfies Equation (1):
\begin{align}
    I = YV
\end{align}
where $I_j \in \mathds{C}^k$ represents a subset of entries of the vector $I$ corresponding to bus $j$. The dimension of $I_j$ can vary between 1 and 3, depending on the number of phases associated with bus $j$. Bus $1$ is chosen to be the reference bus and contains three phases. The voltage magnitude of the nodes in the reference bus is fixed at $1$ and the phase angles are separated by $120^\circ$.
The following subsection details some of the typical measurements that are usually available on the distribution network.
\subsection{Measurements}
The distribution network is typically instrumented with different types of metering equipment serving various purposes (smart meters for billing, sensors associated with telemetered protection equipment, etc.) and these can act as a source of measurement data for state estimation purposes. The mathematical relationship between the voltage phasor $V$ and the different types of  potentially available measurements is given by the \textit{measurement functions} below:

\begin{align}
h_{i_{i \rightarrow j}}(V) &= (V_i - V_j) Y_{ij}\\
h_{|I_{i \rightarrow j}|}(V) &= |h_{I_{i \rightarrow j}}(V)|\\
h_{I_i}(V) &= (YV)_i\\
h_{|V_i|}(V) &= |V_i|\\
h_{S_i}(V) &= (V \circ (YV)^{\star})_i
\end{align}
where the voltage phasor $V$, $h_{i_{i \rightarrow j}}(V)$ is the current flow along the branch $(i,j)$, $h_{I_i}(V)$ is current injection into bus $i$, $h_{|V_i|}(V)$ is the voltage magnitude at bus $i$, $h_{|I_{i \rightarrow j}|}(V)$ is the branch current magnitude, and $h_{S_i}(V)$ is the apparent power injection into bus $i$. Note that $(.)^{\star}$ denotes the complex conjugate and $\circ$ represents the pointwise product of vectors.

Let $\Sigma_m$, where $m \in \{i_{a \rightarrow b},~I_a,~|V_a|, |I_{a \rightarrow b}|, S_a \}$, denote the error covariance associated with each of the measurements. The variances associated with the real-time sensor measurements are typically small and are taken to be constant for the sake of simplicity. The load injection into each bus $S_a$ is usually not available on all buses and will be estimated  as \textit{pseudo-measurement} from historical data (as detailed in Section \ref{sec:4}). 

For a given distribution network with a fixed number $N_m$ of sensors, it is possible to construct a \textit{composite} measurement function $H: \mathds{C}^M \rightarrow \mathds{C}^{N_m}$ that maps the voltage $V$ into the corresponding set of measurements. Similarly, a composite covariance matrix $\Sigma_{meas} \in \mathds{C}^{N_m} \times \mathds{C}^{N_m}$ can be derived by constructing a block diagonal matrix in which the diagonal entries correspond to the covariance matrices associated with the individual measurements.

\section{DSSE Problem Formulation}\label{sec:3}
Given the $composite$ measurement function detailed in the previous section, the DSSE problem with measurements can be formulated as a nonlinear WLS problem as follows: 


\begin{align}
\label{eqn:dssevi}
\hat{V} = &\underset{V}{\argmin}~(H(V) - z)^T\Sigma_{meas}^{-1}(H(V) - z)\\
&s.t.~v_{min} \leq|V_i| \leq v_{max}, -\pi \leq \angle{V_i} \leq \pi~\forall i \in \{1,2,\dots N\}
\nonumber
\end{align}

where $H$ is the composite measurement function, $z \in \mathds{C}^{N_m}$ is the observed measurement and  $\Sigma_{meas}$ represents the error covariance associated with the measurements. The voltage magnitude is constrained to be between $v_{min}$ and $v_{max}$. The upper bound for the voltage magnitudes is chosen to be 1.1, as the load tap changers can bump the voltage as high as 1.05 and the PV injections can further exacerbate the situation. \\
The problem described by Equation (\ref{eqn:dssevi}) is nonconvex due to the magnitude measurements and the apparent power \textit{pseudo-measurements}. An implementation of the optimization problem was done in Julia and MATLAB using the general nonlinear program solver IPOPT.

\section{Scenario Construction and Pseudo-Measurements}\label{sec:4}
\begin{figure}[t]
\centering
\includegraphics[scale=0.4]{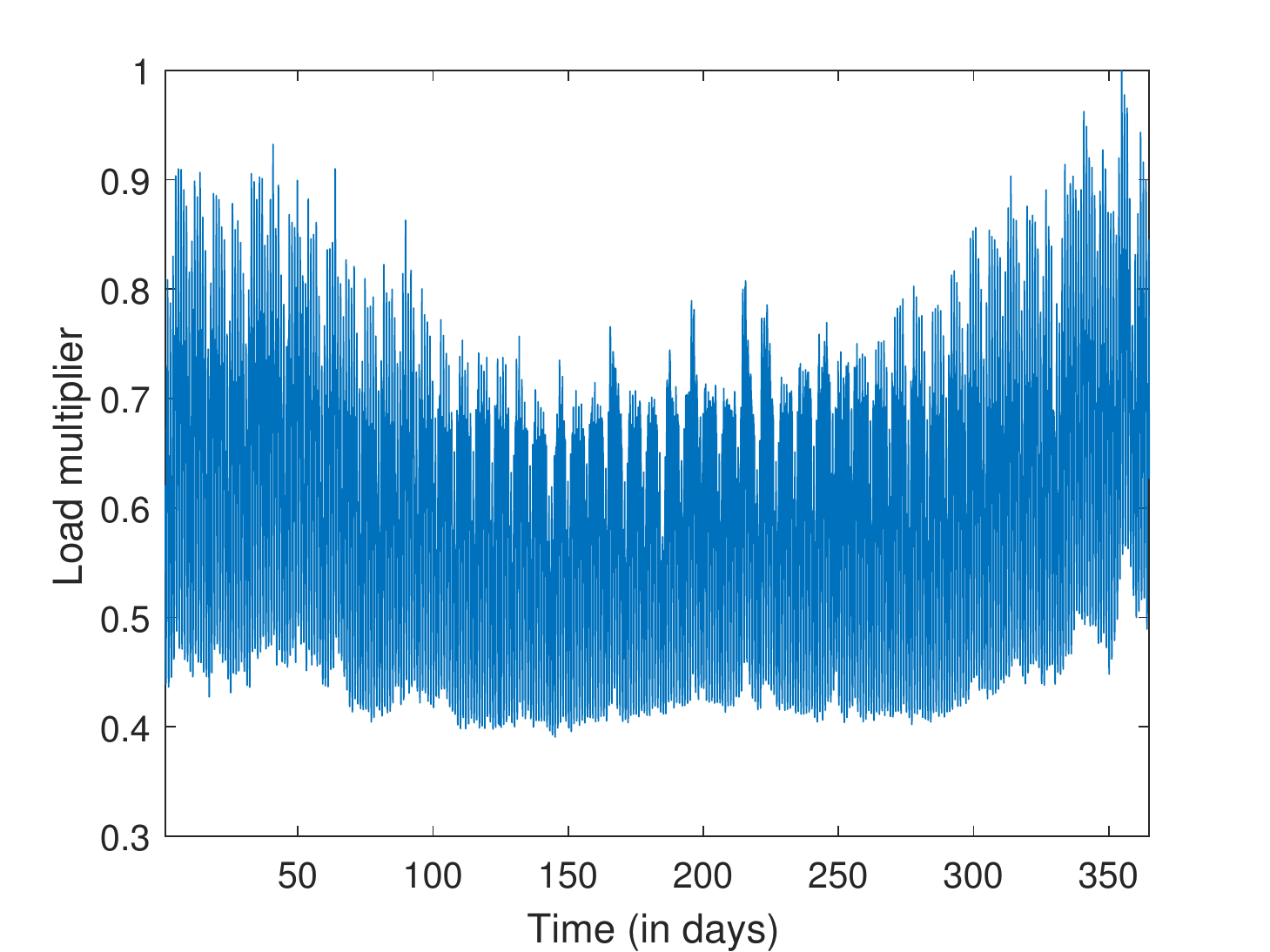}
\caption{Yearlong hourly load multiplier data set}
\label{fig:yearlongL}
\end{figure}

\begin{figure}[t]

\centering
\includegraphics[scale=0.4]{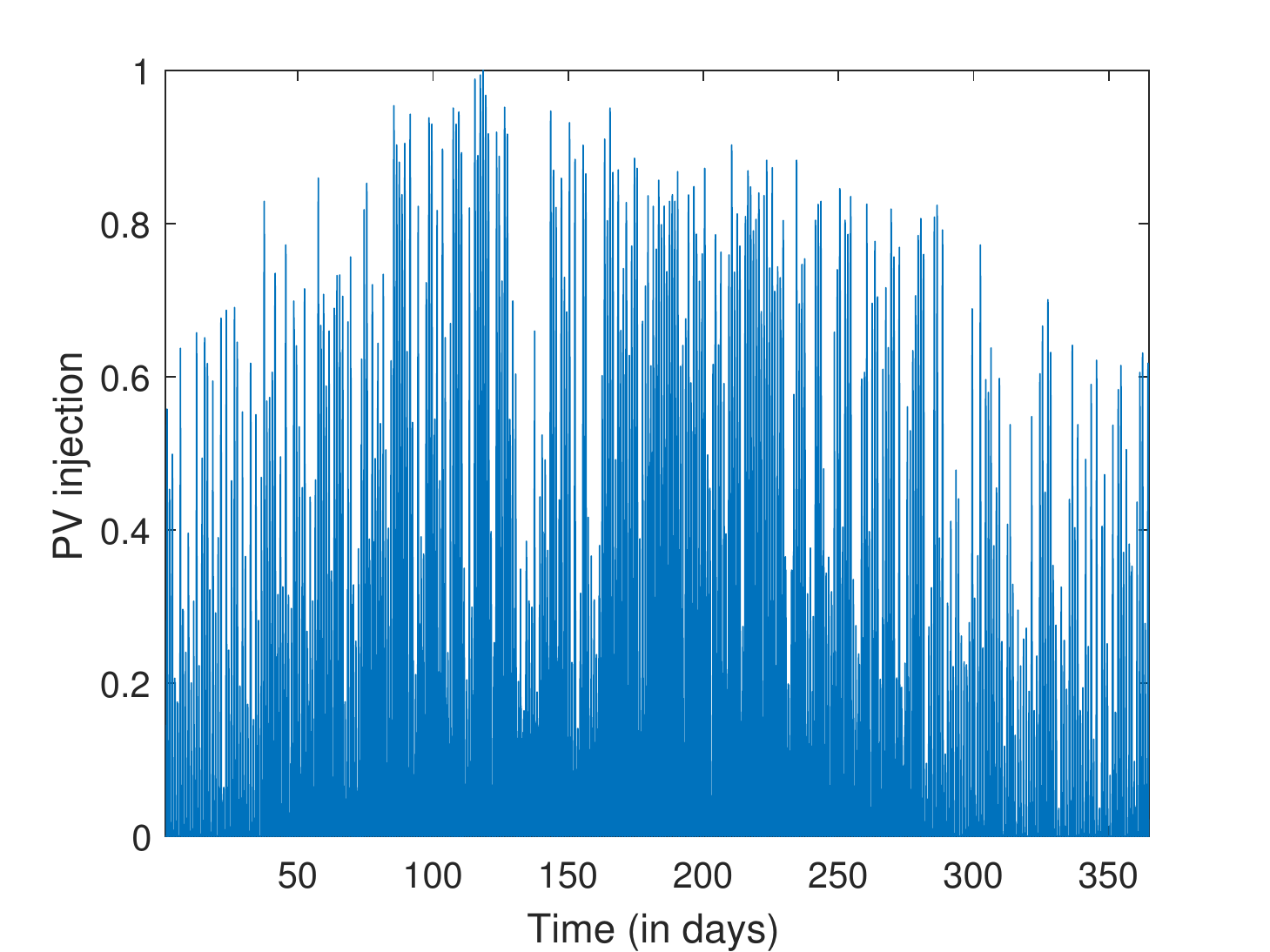}
\caption{Yearlong hourly data set for PV injection}
\label{fig:pv}
\end{figure}

An hourly load multiplier data set $\mathcal{L}$ (shown in \figref{fig:yearlongL} and \figref{fig:pv}), obtained from the OpenDSS simulation platform, describes how the load at the head of the feeder varies throughout the year. The actual load at any node in the feeder at any point in the year is obtained by scaling the nominal load value for that node and scaling it by the load multiplier value for that time of year. Note that the load multiplier data set is synthetic and is only used for generating different scenarios to determine the efficacy of the state estimator.

Similarly, PV forecast/measured data (downsampled from a data set with a resolution of 5 minutes) from  Hinesburg, Vermont, USA, obtained from the National Renewable Energy Laboratory at \url{https://www.nrel.gov/grid/solar-power-data.html}, was normalized to make it compatible with the load multiplier data set $\mathcal{L}$ (shown in \figref{fig:pv}).
 
 Let $\bar{S} = \bar{P} + j\bar{Q} \in \mathds{C}^M$, where $M$ is the number of nodes, denote the nominal real and reactive injections into each of the nodes. Given the solar data $\mathcal{S}$, the load multipliers $\mathcal{L}$, and the nominal load $\bar{S}$, the $k$-th scenario is constructed as follows:
 \begin{align*}
 &P^k = (\alpha_k - s_k) \bar{P} + \epsilon_k \circ \bar{P} \\
 &Q^k = \alpha_k\bar{Q} + \epsilon_k \circ \bar{Q}\\
 &\alpha_k \in\mathcal{L},~s_k \in \mathcal{S}\\
 &\epsilon_k \sim Uniform([-c \alpha_k, c \alpha_k]) \in \mathds{R}^N
 \end{align*}
The parameters $\alpha_k$ and $s_k$ represent the contributions of the base load and the solar generation, respectively, for the k-th scenario. The operator $\circ$ represents the element-wise product of vectors. The vector $\epsilon_k$ is sampled uniformly from the interval $[-c \alpha_k, c \alpha_k]$ and represents small uncertainties in the underlying load profile. The solar injection does not affect the reactive power injection. Since both the solar data $\mathcal{S}$ and the load multiplier data $\mathcal{L}$ represent hourly data for an entire year, 8760 ($365 \times 24$) scenarios are generated. 

\begin{figure}[t!]
\centering
\includegraphics[scale=0.6]{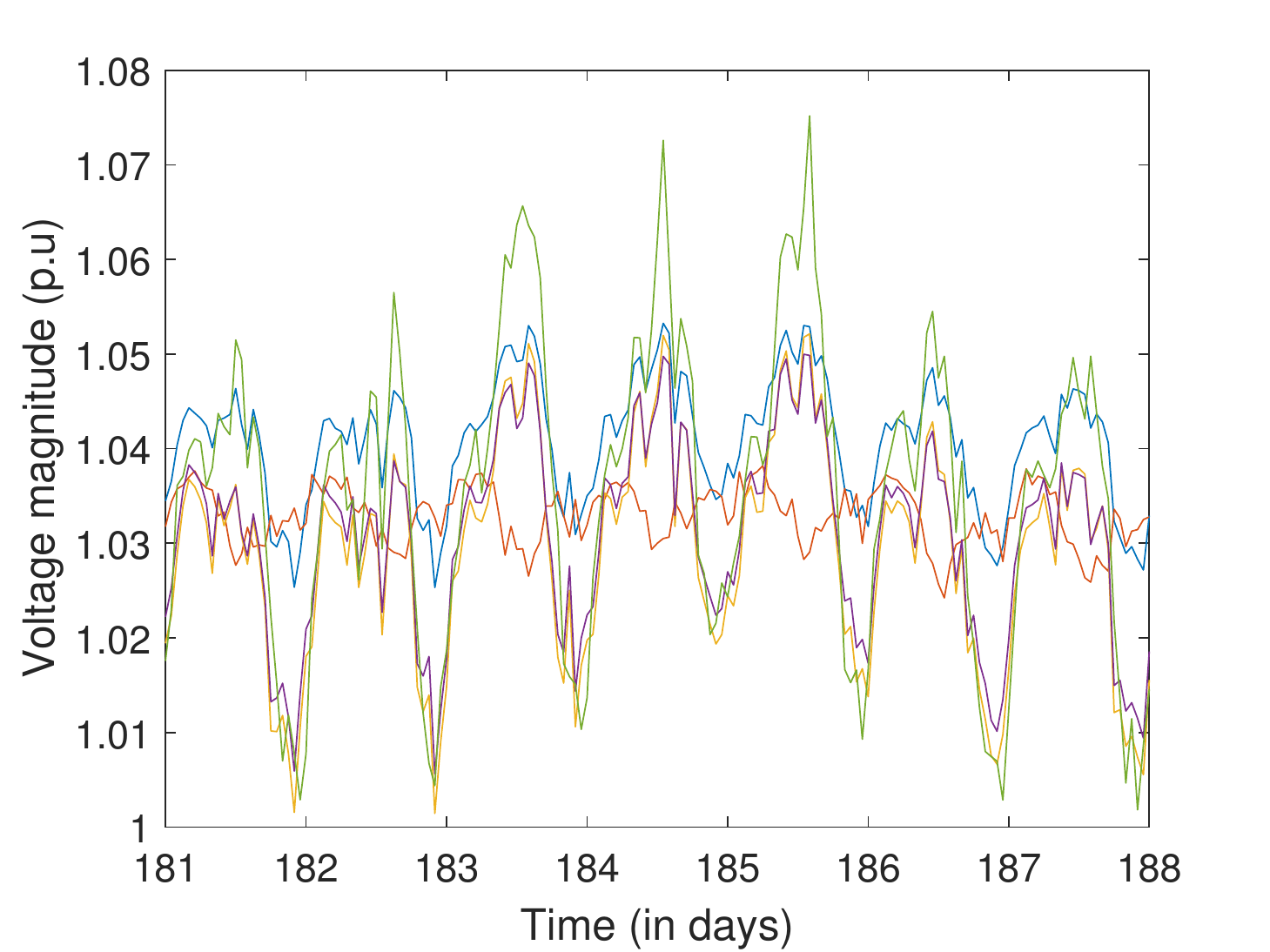}
\caption{Voltage measurements on a subset of the IEEE 13 bus system for a week in summer}
\label{fig:vprofiles}
\end{figure}

The voltage profile $V^k$ corresponding to each scenario $k$ is generated by solving the load flow equation in OpenDSS (for the IEEE 13 bus system) and MATPOWER (for the IEEE 33 bus system). A small subset of the generated voltage profiles (for the 13 bus system) is shown in \figref{fig:vprofiles}. Note that the tap changers are fixed at their full load position, increasing the voltage magnitude to 1.05 p.u. at the head of the feeder. As such, all the voltage magnitudes are above 1.0 p.u. The dip observed in the voltage profile corresponds to nighttime, when the PV injection is minimal, while the peaks correspond to daytime, when there is surplus power, due to the PV injections, fed back into the power grid.

The \textit{pseudo-measurements} for the real and reactive injections are generated by taking the mean and the variance of the load multiplier and the solar data as follows:
\begin{align*}
&\hat{P} = \dfrac{1}{|\mathcal{L}|} \sum_{k=1}^{|\mathcal{L}|} P^k\\
&\hat{Q} = \dfrac{1}{|\mathcal{L}|} \sum_{k=1}^{\mathcal{L}} Q^k\\
&\Sigma_P = \dfrac{1}{|\mathcal{L}|-1} \sum_{k=1}^{|\mathcal{L}|} (\hat{P} - P^k) (\hat{P}-P^k)^T\\
&\Sigma_Q= \dfrac{1}{|\mathcal{L}|-1} \sum_{k=1}^{|\mathcal{L}|} (\hat{Q} - Q^k) (\hat{Q}-Q^k)^T
\end{align*}

\subsection{Error Metrics}
The error metric used for evaluating the state estimator is the percentage error relative to the actual voltage magnitude. If $v \in \mathds{R}^N$ is the voltage magnitude at each of the nodes and $\hat{v} \in \mathds{R}^N$ is the voltage estimate constructed by the state estimator, then 
\begin{align*}
\%{NodeError}_i = \dfrac{|v_i - \hat{v}_i|}{|v_i|} \times 100
\end{align*}
where $v_i$, $\hat{v}_i$ represents the voltage magnitude at node $i$, and $\hat{v}_i$ is the voltage estimate at node $i$.

\begin{figure}
\centering
\includegraphics[scale=0.5]{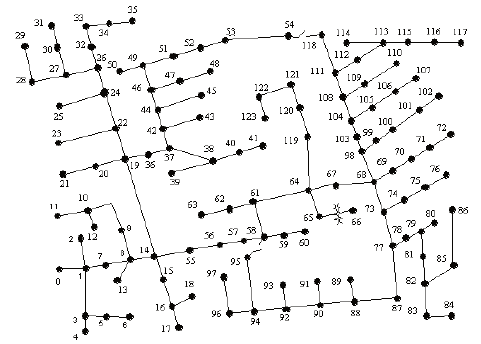}
\caption{IEEE 123 bus feeder configuration}
\label{fig:ieee123}
\end{figure}
\section{Numerical Results}\label{sec:5}
The DSSE problem (\ref{eqn:dssevi}) was solved in MATLAB/Julia using IPOPT, and was tested in three different test systems: a balanced 33 bus test feeder, an unbalanced IEEE 13 bus test system (shown in \figref{fig:13bus}), and an unbalanced IEEE 123 bus test system (shown in \figref{fig:ieee123}). 
\subsection{State Estimator Performance}
\begin{figure}[t]
\centering
\includegraphics[scale=0.6]{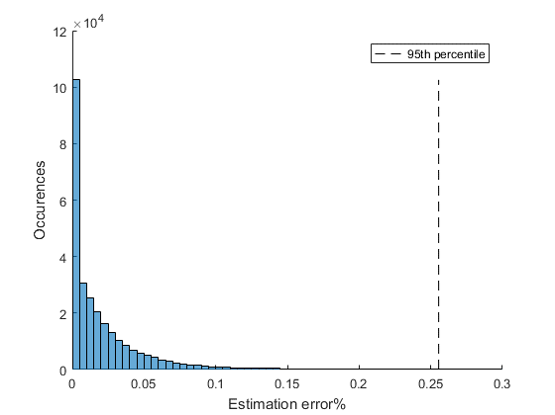}
\caption{95th percentile of aggregate $\%NodeError_i$ for the 33 bus system} 
\label{fig:33busAgg}
\end{figure}

\textit{MATPOWER 33 bus test feeder:} For the balanced MATPOWER 33 bus test system, voltage and phase angle measurements are taken from buses 8, 9, 12, and 25. Bus 1 is fixed at 1.0 p.u. and is used as the reference bus. \figref{fig:33busAgg} shows the 95th percentile of the aggregate $\%{NodeError}_i$ (i.e., the node error across all the buses).

\textit{IEEE 13 bus feeder:} The 13 bus system is unbalanced and has 41 nodes, as each bus has multiple nodes corresponding to different phases. Voltage measurements are taken from nodes $10$, $11$, and $12$ (corresponding to three phases of bus 633), $24$ (corresponding to phase 2 of bus 670), and $29$ (corresponding to phase 1 of bus 680). Nodes 1 to 3, corresponding to the source bus, are fixed at 1.0 p.u. and are used as the reference bus.

\begin{figure}[t]
\centering
\includegraphics[scale=0.6]{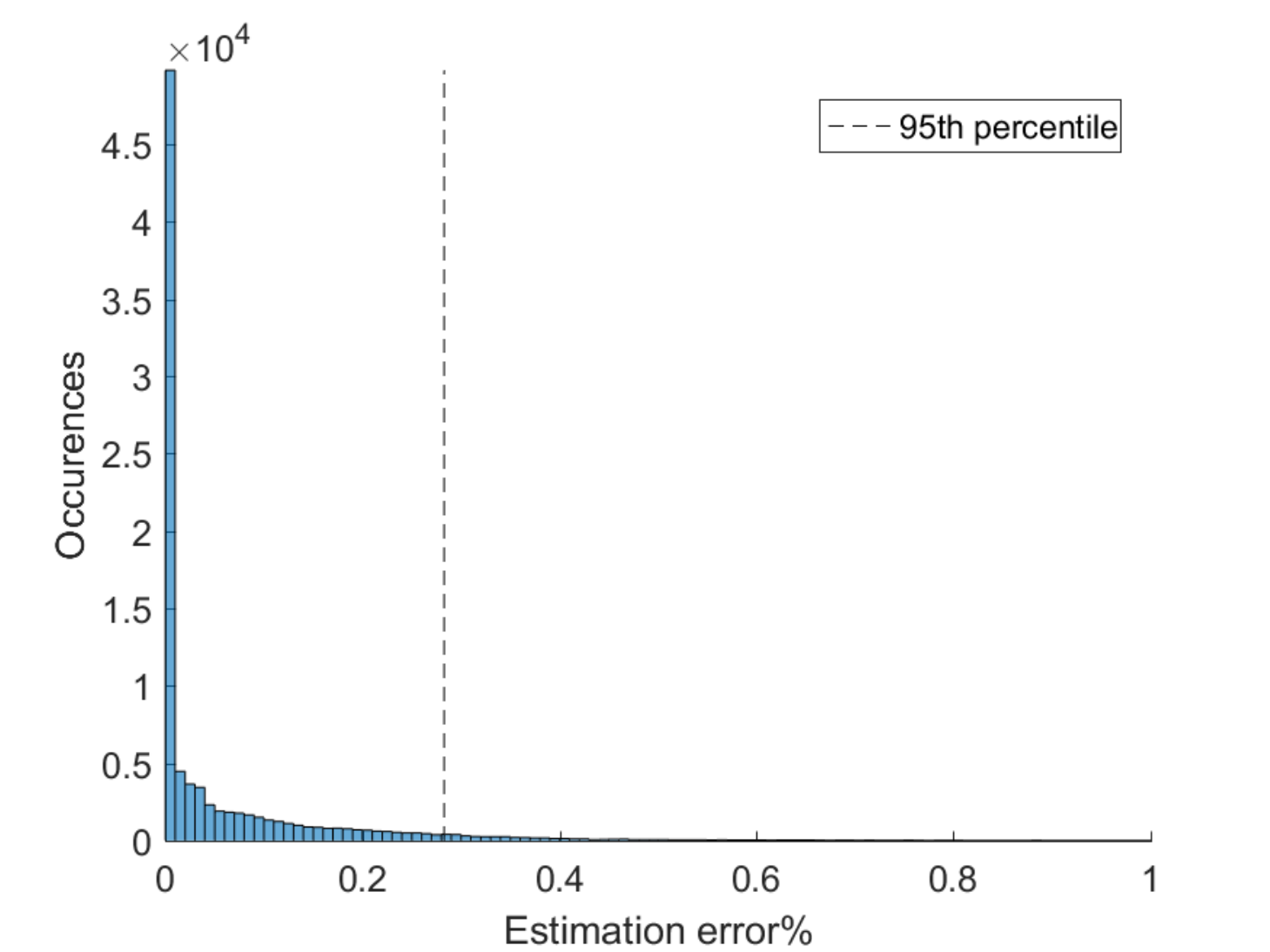}
\caption{95th percentile of aggregate $\%NodeError_i$ for the 13 bus system}
\label{fig:13busAgg}
\end{figure}

Like that for the 33 bus system, \figref{fig:13busAgg} plots the value of the 95th percentile of the quantity $\%{NodeError}_i$ for each bus and in aggregate for the summer months.

\begin{figure}[t]
\centering
\includegraphics[scale=0.6]{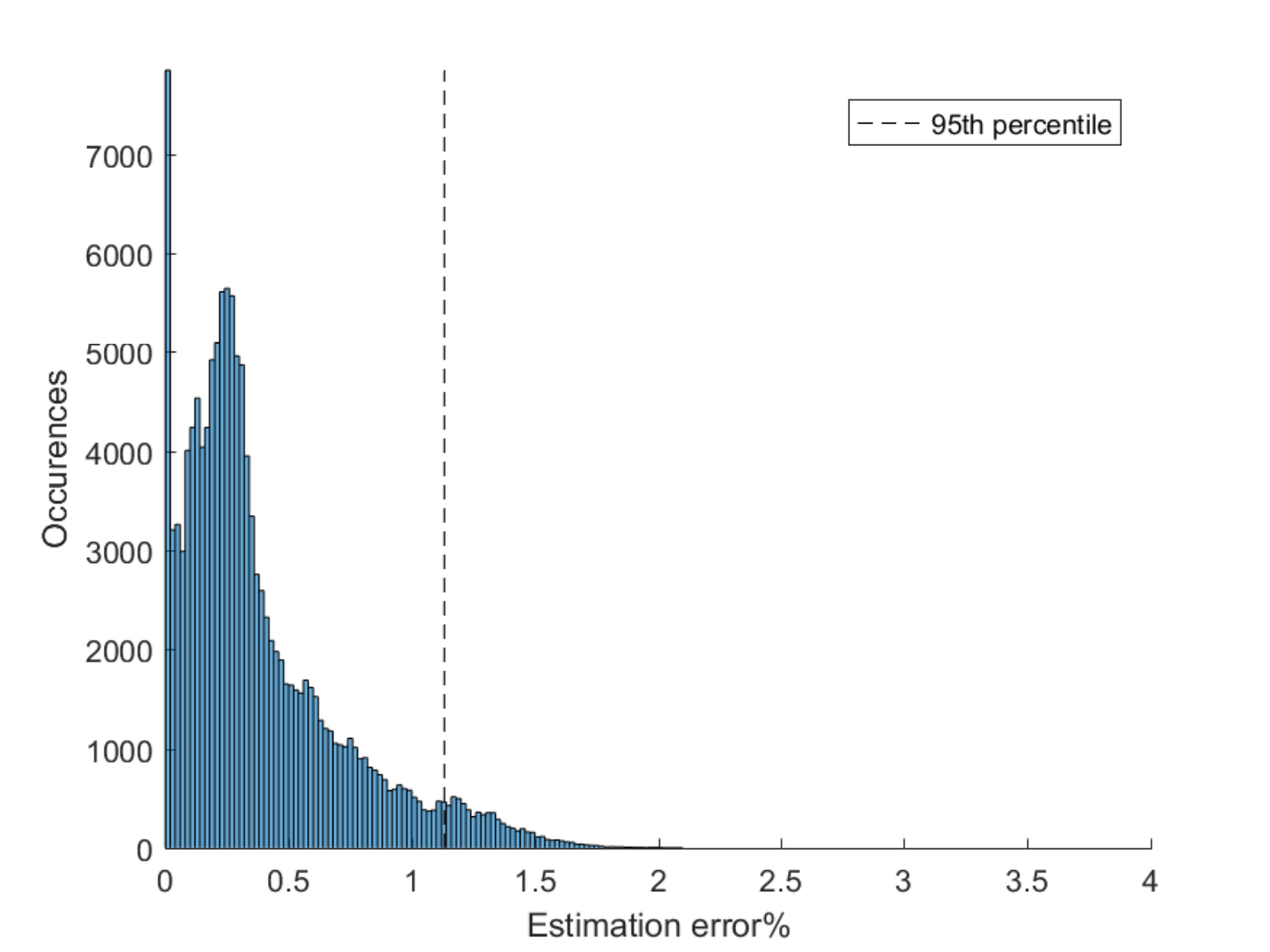}
\caption{95th percentile of aggregate $\%NodeError_i$ for the 123 bus system}
\label{fig:123busAgg}
\end{figure}

\textit{IEEE 123 bus feeder} The IEEE 123 bus system has 278 nodes, the majority of which (roughly 68\%) are unloaded. \figref{fig:123busAgg} shows the value of the 95th percentile of the quantity $\%{NodeError}_i$ in aggregate, for a summer month, where voltage measurements are taken from 20 different nodes.

It can be seen in that all three cases, the overall error is less than 2\% for 95\% of the test scenarios.

\subsection{Sensitivity to Sample Variance}
\begin{figure}[t]
\centering
\includegraphics[scale=0.6]{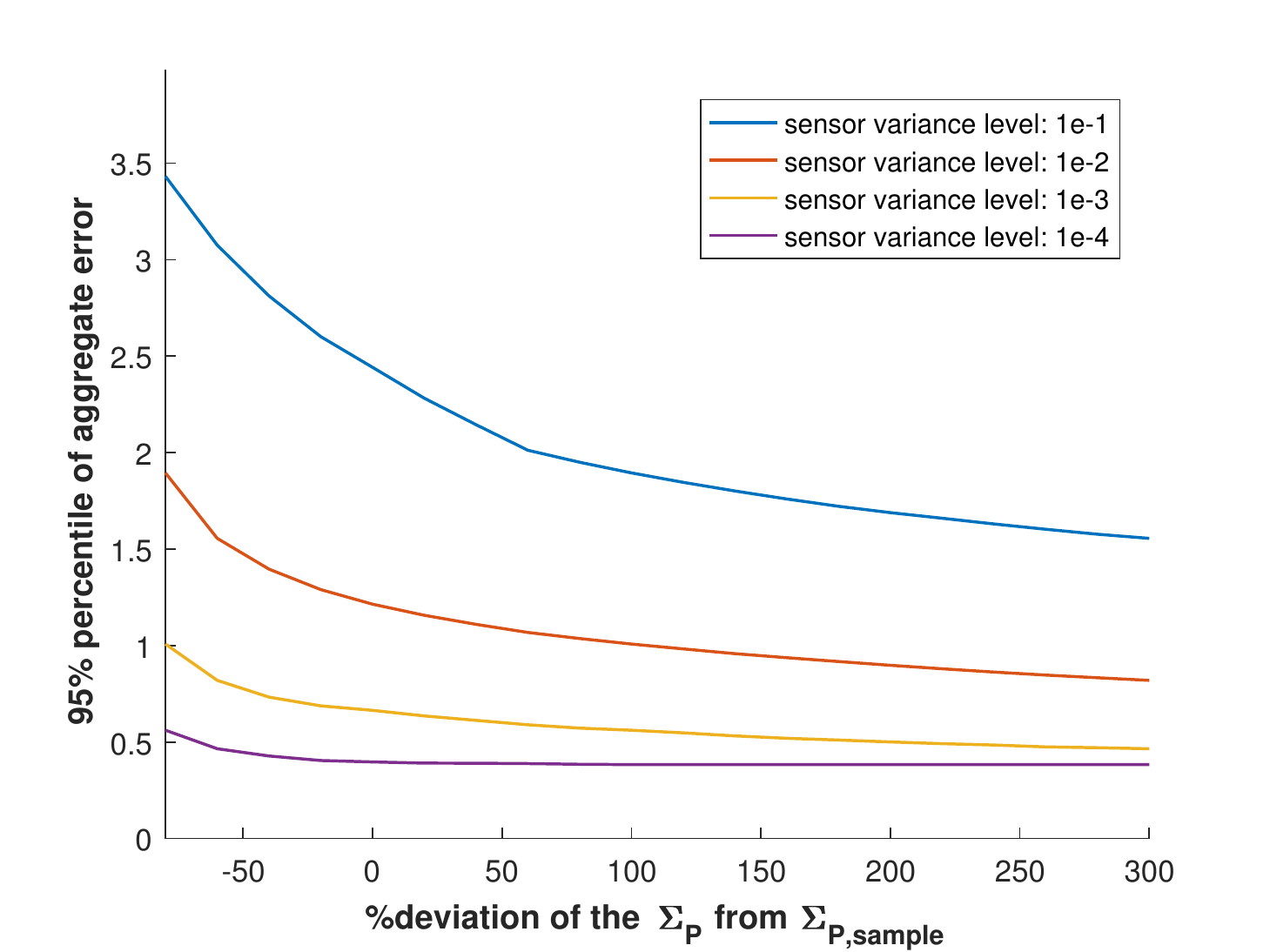}
\caption{95th percentile of aggregate $\%NodeError_i$ for the 33 bus system as a function of the deviation of $\Sigma_P$ from $\Sigma_{P,sample}$ at varying sensor noise levels}
\label{fig:prctile33}
\end{figure}

\begin{figure}[t]
\centering
\includegraphics[scale=0.6]{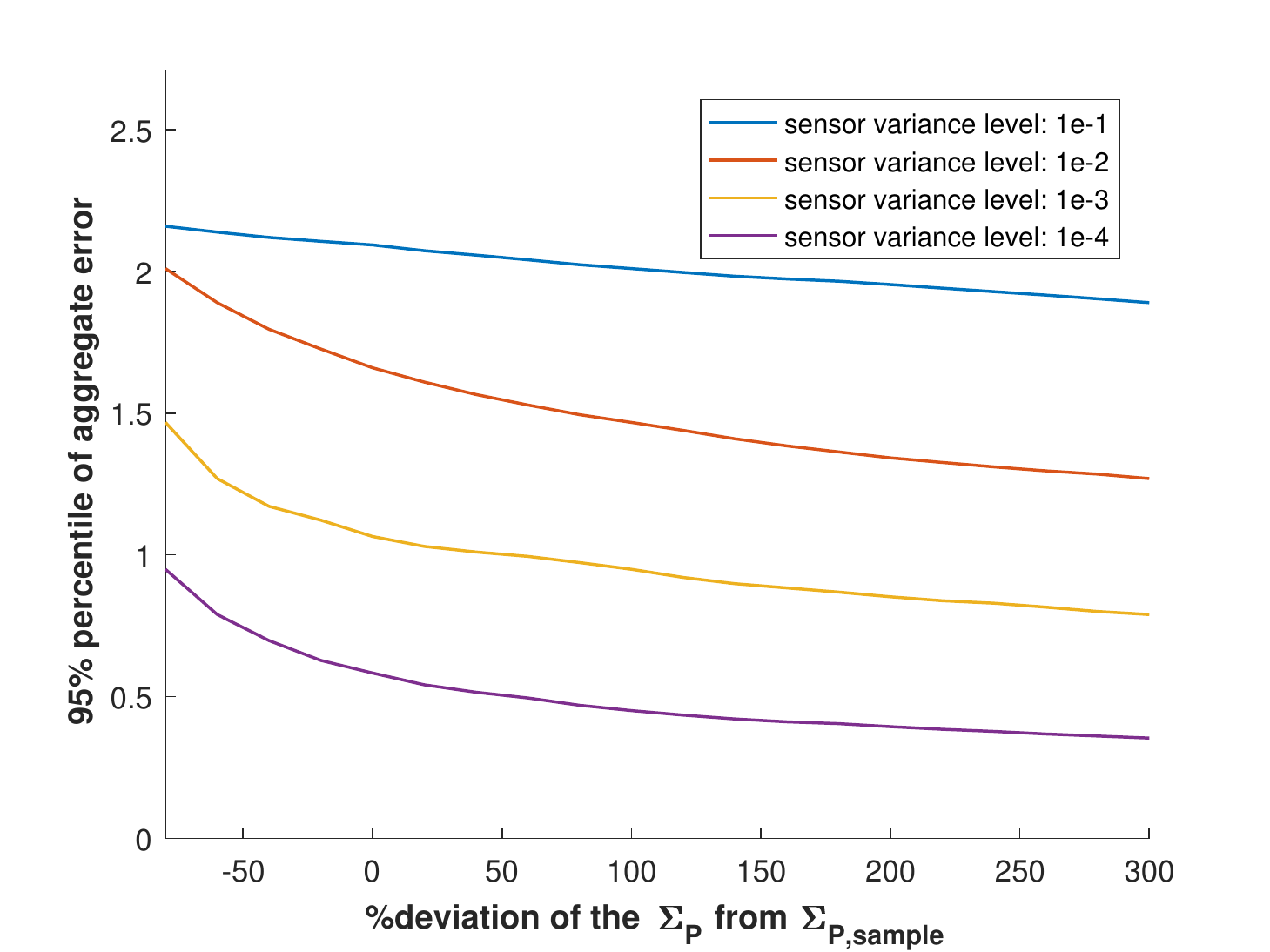}
\caption{95th percentile of aggregate $\%NodeError_i$ for the 13 bus system as a function of the deviation of $\Sigma_P$ from $\Sigma_{P,sample}$ at varying sensor noise levels}
\label{fig:prctile13}
\end{figure}

In typical situations, the sample mean and variance  of the pseudo-measurements are obtained from historical data. As such, it is likely that the sample variance, $\Sigma_{P, sample}$ and $\Sigma_{Q, sample}$, associated with a particular time period (say, a week in summer) differs from the historical variance, $\Sigma_P$ and $\Sigma_Q$ used for the pseudo-measurements. Thus, it is of interest to understand the behavior of the state estimation algorithm when the variance of the pseudo-measurement deviates from the actual sample variance. \figref{fig:prctile33} and \figref{fig:prctile13} show the 95th percentile of the aggregate $NodeError_i$ 
(for a summer week) as a function of the percent deviation of $\Sigma_P$ from $\Sigma_{P,sample}$ ($\Sigma_{Q}$ is perturbed in a similar way) at various sensor error covariance levels. Note that a reduction in the overall sensor noise reduces the overall level of error in the estimates, while the percentile error exhibits a monotone decrease as the pseudo-measurement is increased. Underestimating the variance of the pseudo-measurements relative to that of the actual sample variance (i.e., considering the pseudo-measurements to be more accurate) results in larger error because it is likely that the pseudo-measurements contradict the sensor readings, which are far more reliable. As such, increasing the variance of the pseudo-measurement decreases the overall percentile error.


\begin{figure}[t]
\centering
\includegraphics[scale=0.6]{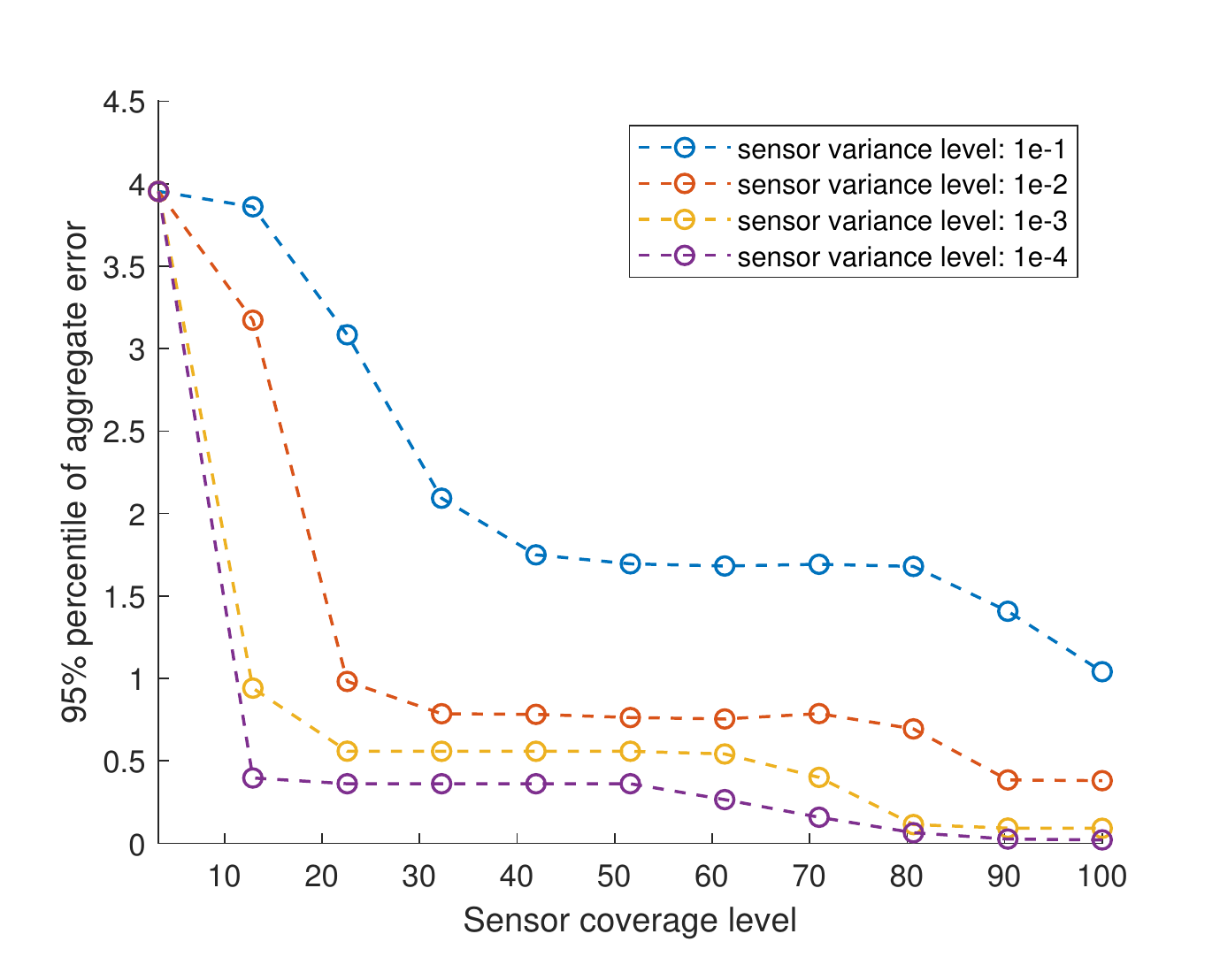}
\caption{95th percentile of aggregate $\%NodeError_i$ for the 33 bus system as a function of different sensor coverage levels at varying sensor noise levels}
\label{fig:cov33}
\end{figure}

\begin{figure}[t]
\centering
\includegraphics[scale=0.6]{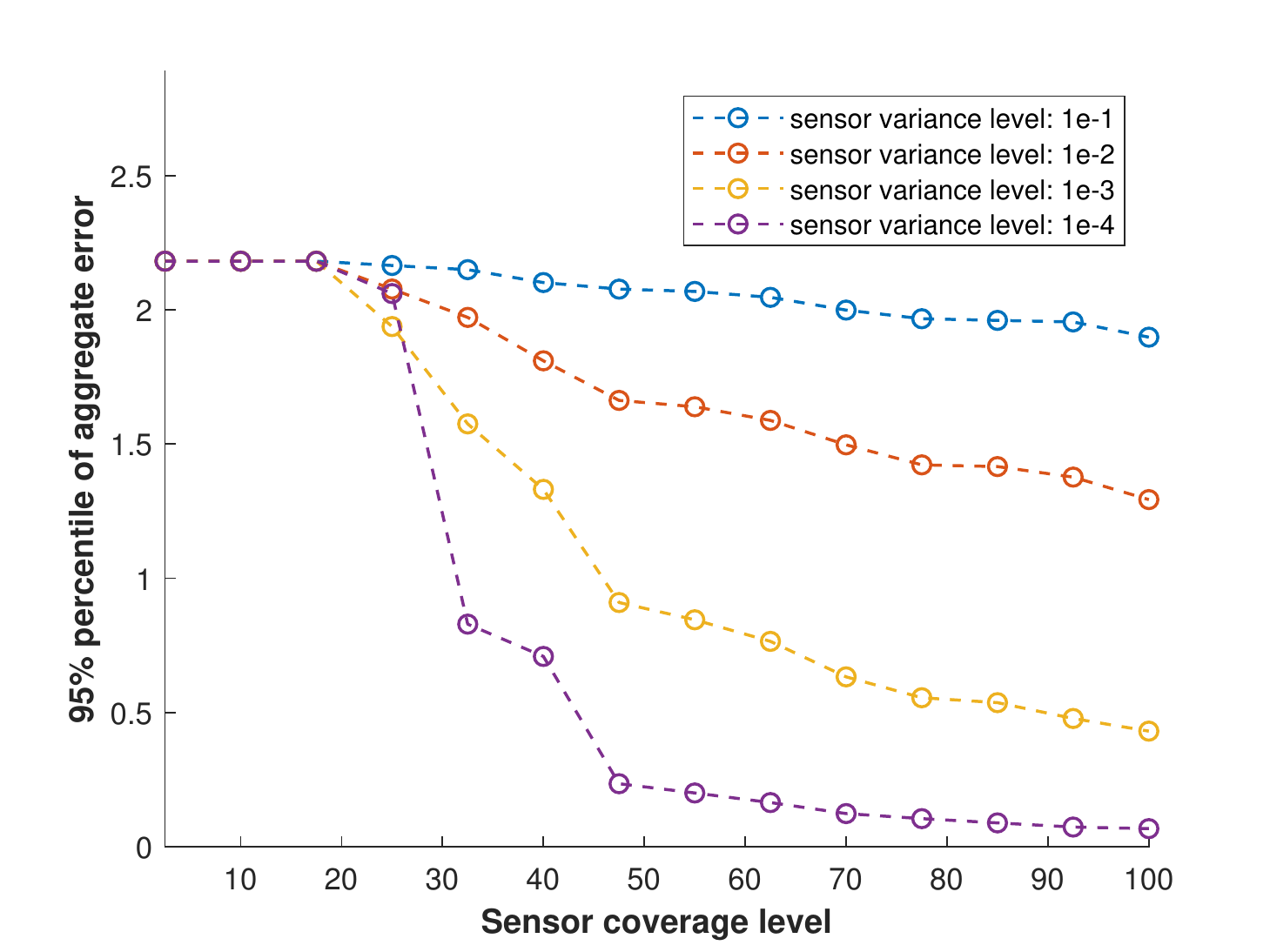}
\caption{95th percentile of aggregate $\%NodeError_i$ for the 13 bus system as a function of different sensor coverage levels at varying sensor noise levels}
\label{fig:cov13}
\end{figure}

\subsection{Sensor Coverage/Noise}
To minimize the cost of sensor deployment, it is important to identify key locations where placing a sensor would improve the quality of the state estimate. \figref{fig:cov33} and \figref{fig:cov13} show how the 95th percentile of aggregate $\%NodeError_i$ (for a summer week) varies as the sensor coverage is increased (at different levels of sensor error, as indicated by the increasing sensor error covariance) for the 13 bus and 33 bus feeders. The sensors were added sequentially (three at a time for the 13 bus system, two at a time for the 33 bus feeder), starting at the head of the feeder. In both cases, it is interesting to note that there is a large drop in the aggregate error when sensors are added to certain locations (corresponding to node 632 for the 13 bus feeder, and corresponding to node 2 in the 33 bus feeder, both of which represent points at which the feeder branches out into several trunks). Furthermore, it can also be seen that the error remains relatively flat (especially for the 13 bus feeder), implying only a few sensors are required to allow adequate estimation of the state of the feeder.

\section{Concluding Remarks}
In this paper, a distribution system state estimation problem was formulated, and a scenario generation framework was proposed for testing the state estimator under a wide variety of conditions. The sensitivity of the state estimator accuracy to sensor accuracy and sensor coverage levels via simulation was also investigated.

\section{Acknowledgements}
This work was performed for the U.S. Department of Energy under Contract DE-AC05-76RL01830 with support from the ENERGISE program and the Grid Modernization Lab Initiative.
\bibliography{dsse-lit} 
\bibliographystyle{ieeetr}
\end{document}